\documentstyle[12pt,a4wide]{article}

\def\lta{~\raise.4ex\hbox{$<$}\llap{\lower.6ex\hbox{$\sim$}}~}
\def\gta{~\raise.4ex\hbox{$>$}\llap{\lower.6ex\hbox{$\sim$}}~}

\begin{document} \input psfig.sty
\begin{titlepage}

\title{Spatial patterns and scale freedom in a Prisoner's Dilemma 
cellular automata with Pavlovian strategies}
\author{H. Fort$^{1}$ and S. Viola$^{2}$}

\maketitle

\nopagebreak

1: Instituto de F\'{\i}sica, Facultad de Ciencias, Universidad de la
Rep\'ublica, Igu\'a 4225, 11400 Montevideo, Uruguay\\

2: Instituto de F\'{\i}sica, Facultad de Ingenier\'{\i}a, Universidad de la
Rep\'ublica, Julio Herrera y Reissig 565, 11300 Montevideo, Uruguay. Currently at Department of Physics, Boston University, 590 Commonwealth Avenue, Boston, MA 02215, USA.

\nopagebreak 

\begin{abstract}
 
A cellular automaton in which cells represent agents playing the
Prisoner's Dilemma (PD) game following the simple "win-stay, 
loose-shift" strategy is studied.
Individuals with binary behavior, such as they can either cooperate (C) or defect
(D), play repeatedly with their neighbors (Von Neumann's and Moore's 
neighborhoods).
Their utilities in each round of the game are given by a rescaled payoff 
matrix described by a single parameter $\tau$, which measures the ratio
of {\it temptation to defect} to {\it reward for cooperation}.
Depending on the region of the parameter space $\tau$, the system 
self-organizes - after a transient -  into dynamical equilibrium states
characterized by different definite fractions of C agents $\bar{c}_\infty$ 
(2 states for the Von Neumann neighborhood and 4 for Moore neighborhood). 
For some ranges of $\tau$ the cluster size distributions, the
power spectrums $P(f)$ and the perimeter-area curves follow power-law scalings.
Percolation below threshold is also found for D agent clusters.
We also analyze the asynchronous dynamics version of this model and
compare results.
 
\end{abstract}

\vspace{2mm}
 
{\it keybords}: {\it Complex adaptive systems, Sociophysics, Econophysics,
Agent-based models, Self-organized criticality.}
 
\vspace{1mm}
 

\end{titlepage}

\section{Introduction}

The {\it Prisoner's Dilemma} (PD) game plays in Game Theory a role
similar to the harmonic oscillator in Physics.
Indeed, this game, developed in the early fifties, offers a very simple and
intuitive approach to the problem of how cooperation emerges in societies of
"selfish" individuals {\it i.e.} individuals which pursue exclusively their
own self-benefit. It was used in a series of works by Robert Axelrod and
co-workers \cite{axel84} to examine the basis of cooperation
in a wide variety of contexts.
Furthermore, approaches to cooperation based on the PD have shown their
usefulness in Political Science \cite{h90}-\cite{g88},
Economics \cite{ff02}-\cite{w89}, International Affairs  \cite{h01}-\cite{s71},
Theoretical Biology \cite{wn99}-\cite{n90} and Ecology
\cite{md97}-\cite{dmh92}.

The PD game consists in two players each confronting two
choices: cooperate (C) or defect (D) and each makes his choice without
knowing what the other will do.
The four possible outcomes for the interaction of both agents 
are: 1) they can both cooperate: (C,C), 2) both defect: (D,D), 3) 
one of them cooperate and the other defect: (C,D) or (D,C).
Depending on the case 1)-3), the agents get respectively
: the "reward" $R$, the "punishment" $P$ or the "sucker's
payoff" $S$ the agent who plays C and the "temptation to defect" $T$ 
the agent who plays D.
These four payoffs obey the
relations: 
\begin{eqnarray}
T>R>P>S \nonumber \\ 
\mbox{and} \nonumber \\
2R>S+T. 
\label{eq:ine}
\end{eqnarray}

The last condition is
required in order that the average utilities for each agent of a cooperative
pair ($R$) are greater than the average utilities for a pair exploitative-
exploiter (($R+S$)/2).
One can assign a {\em payoff matrix} M to the PD game given by
\vspace{-4mm}
 
\begin{center}
 
$${\mbox M}=\left(\matrix{(R,R)&(S,T)\cr (T,S)&(P,P) \cr}\right),$$
 
\end{center}
which summarizes the payoffs for {\it row} actions when confronting
with {\it column} actions.
Clearly it pays more to defect: if one of the two players defects -say $i$-
, the other who cooperates will end up with nothing. In fact, even if agent
$i$ cooperates, agent $j$ should defect, because in that case he will get $T$
which is larger than $R$.
That is, independently of what the other player does,
defection D yields a higher payoff than cooperation and is the
{\it dominant strategy} for rational agents.
Furthermore, is the {\it Nash equilibrium} 
\cite{n51} - {\it i.e.} a best reply to itself - of the PD game.
The dilemma is that if both defect, both do worse than if
both had cooperated: both players get $P$ which is smaller than $R$.
A possible way out for this dilemma is to play the game repeatedly.
In this iterated Prisoner's Dilemma (IPD), there are several strategies
that outperform the dominant [D,D] one-shot strategy
and lead to some non-null degree of cooperation.
 
The attainment of cooperation in PD simulations relies on different
mechanisms and factors.
A popular point of view regards direct reciprocity as the crucial
ingredient. A typical exponent of this viewpoint is the strategy known
as {\it Tit for Tat} (TFT): cooperate on the first
move, and then cooperate or defect exactly as your opponent did on
the preceding encounter.
This requires either memory of previous interactions or features ("tags")
permitting cooperators and defectors to distinguish one another
\cite{ep98}.
 
Spatial structure has also been identified as an influential factor in
building cooperation. For instance, in ref. \cite{nm92}
the authors neglected
all strategical complexities or memories of past encounters. Instead, they
show that spatial effects by themselves in a classic Darwinian setting
are sufficient for the evolution of cooperation
\footnote{Indeed the game they considered is not exactly the PD and
implies a "weak dilemma" in which D does not strictly
dominate.}.

The problem of cooperation is approached
mainly from an Darwinian evolutionary perspective: strategies that
incorporate some dose of cooperative behavior are the most successful
and propagate displacing competing strategies that do not.
In that sense, a central concept is that of {\it evolutionary stable
strategy} (ESS) \cite{msp73}, \cite{ms82}: a strategy which if adopted
by all members of a population cannot be invaded by a mutant strategy
through the operation of natural selection.
The {\it evolutionary game theory}, originated as an application of
the mathematical theory of games to biological issues, later
spread to economics and social sciences.
 
In this work, we follow a different approach: there is no
competition of different strategies, all the agents follow
a natural strategy of "win-stay, loose-shift" known as Pavlov
\cite{kk88}.
We do not worry about the resistance of the strategy against invasion by
other strategies (like unconditional defectors or ALL D that play D
independently of what the opponent does), rather we take Pavlov
for granted. The rationale for this relies on several facts. First, 
Pavlov seems to be a widespread strategy in nature \cite{db86}.
Second, Pavlov does pretty well when competing
with several other strategies including {\it generous tit-for-tat} GTFT
\footnote{GTFT cooperates after the opponent has cooperated in the previous
round, but it also cooperates with a non null probability after the opponent
has defected.} 
as it was shown by Nowak and Sigmund \cite{ns93}. Moreover, they  
found that in a non-spatial setting while Pavlov can be invaded by ALL D
a slightly stochastic variant cannot.
Third, experiments with humans have shown that a great fraction of
individuals indeed use Pavlovian strategies \cite{mw96}. 

Therefore, we address the analysis of the self-organized states 
that emerge when simple agents, possessing neither long term memory 
nor tags, play the PD game in a spatial setting using Pavlov strategy.
We this aim we resort to a cellular automaton in which each cell is 
either black or 
white representing, respectively, a D or a C agent. Each agent plays with 
those belonging to his neighborhood, and the total utilities he gets determine 
the update of his individual state.

We consider payoff matrices implying strict dilemmas defined by 
equations (\ref{eq:ine}) rather than weak ones in which the inequalities are
relaxed (for instance $P=S$). 
To simplify things we parameterize the payoff matrix in terms of a single
parameter $\tau$, which measures the ratio
of {\it temptation to defect} to {\it reward for cooperation}.

Different self-organizations occur depending on the value of $\tau$, the
type of dynamics and the considered neighborhood. In particular,
for a range of values of $\tau$ (that depends on the neighborhood) we found
power law behavior that might be a signature of 
self-organized criticality \cite{btw87}.
 
Previously, a non spatial similar model, in which 
pairs of agents were chosen at random, was analyzed in ref. \cite{JASSS03}. 
Also, a Mean Field {\it stochastic} version was considered in \cite{PRE04}.

This work is organized as follows. In section 2 we describe the model.
In section 3 we present the results of simulations as well as 
analytical results obtained by using a 
Mean Field approximation that neglects all spatial correlations
(details in the appendix at the end).  
Section 4 is devoted to conclusions and final remarks.

\section{The Model}

The model is vey simple: we assign to each agent, located at
the cell with center at $(x,y)$, a binary behavioral variable $c(x,y)$ which 
takes the value "1" for C agents and "0" for D agents.
This agent plays with the $z$ agents belonging to his neighborhood 
$N(x,y)$ getting a payoff $U_1(x,y)$ with the first neighbor he plays,
$U_2(x,y)$ with the second one and so on \footnote{The order in which a
given agent plays with his neighbors doesn't matter, it can be fixed or 
randomly chosen}.   
The total utilities $U(x,y)=U_1(x,y)+U_2(x,y)+...+U_z(x,y)$ 
he gets playing with his neighborhood
determine the update of his individual state.
More technically, we have an {\em outer totalistic} cellular automaton
{\it i.e.} the state of a cell at the next time-step
depends only on its own state, and the sum of the states of its
neighbors.
The dynamic is synchronous: all the agents update
their states simultaneously at the end of each lattice sweep.
In addition to this synchronous dynamics or ''parallel updating"
we also explored, with less detail, the asynchronous dynamics or ''sequential
updating", in which the
state of an agent is updated after he played.

We considered two different
neighborhoods: a) the {\it von Neumann neighborhood} ($z=4$ neighbor
cells: the cell above and below, right and left from a given cell) and
b) the {\it Moore neighborhood} ($z=8$ neighbor cells: von Neumann
neighborhood + diagonals).

The payoff matrix is parameterized in terms of a single parameter
$\tau \equiv T/R$:

\begin{equation}
{\mbox M}=\left(\matrix{(1,1)&(-\tau,\tau)\cr (\tau,-\tau)&(-1,-1) \cr}\right),
\label{eq:payoffmatrix}
\end{equation}
with $\tau>1$. 
The total utilities of the agent at $(x,y)$ at time $t$, $U(x,y,t)$, are the   
sum of the utilities collected by playing with each of his neighbors, as    
prescribed by the payoff matrix.   

A typical value for the population of agents is $N_{ag}=10,000$
(100 $\times$ 100 lattice)\footnote{However, in some cases we considered 
$N_{ag}$ up to 1,000,000 (1000 $\times$ 1000 lattice) in order the
transients become long enough to extract the power spectrum.}.
 
The initial state at $t=0$ is taken as $c(x,y;0)=0$ or $1$ (D or C 
respectively), chosen at random for each cell $(x,y)$. 
Then the system evolves by iteration during $t_f$ time steps till it reaches
a stationary or dynamical equilibrium state.

Pavlov's strategy works as follows. The agent at $(x,y)$ will change his state 
for the next time step 
$t+1$: $c(x,y,t+1)=1-c(x,y,t)$ (from C to D or viceversa) if 
$U(x,y,t) < 0$, 
and will remain the same: $c(x,y,t+1)=c(x,y,t)$, if $U(x,y,t) > 0$ (when 
$U(x,y,t) = 0$ the agent changes with probability 0.5). 
Once all the agents have played, their state is updated for the next time 
iteration. 

For the von Neumann neighborhood then, each agent plays with his four 
nearest neighbors. Let's analyze what is expected to 
happen for different values of the parameter $\tau$. Let's focus on 
the agent at $(x,y)$ and his possible configurations (C or
D) and the ones of 
his neighborhood (number of C and D neighbors) and in each case 
his corresponding utilities. These results are
shown in Table 1:

\vskip 1mm

\begin{center}

\begin{tabular}{|c||c|c|c|c|c|}  \hline
  & 4C, 0D & 3C, 1D & 2C, 2D & 1C, 3D & 0C, 4D \\ \hline \hline
C   & 4 & 3-$\tau$  & 2-2$\tau$ & 1-3$\tau$ & -4$\tau$  \\ \hline
D   & 4$\tau$ & 3$\tau$ -1  & 2$\tau$-2 & $\tau$-3 & -4  \\ \hline 
\end{tabular}   

\vskip 1mm

Table 1. Utilities of a given agent depending if his state is C (row 1) or 
D (row 2) and the states of his neighborhood (columns 2 to 6) for von
Neumann neighborhood.

\end{center}

From Table 1, since $\tau > 1$, 
we observe that the sign of the utilities $U(x,y)$ of the agent located 
at site $(x,y)$ -which determines the update of his $c(x,y)$- 
depends on the value of $\tau$ only for two cases: a) if the agent plays C 
and his neighborhood consists in 3C agents and 1D or b) if the agent plays D
and his neighborhood consists in 1C agent and 3D agents.
In both cases the update rule depends thus whether $\tau >3$ or $\tau < 3$. 
So, \textit{a priori}, one would expect the existence of a ''critical" value 
of the parameter $\tau^*=3$ such that the results depend on whether $\tau$ is greater
or smaller than this critical value.
Intuitively one can argue that since 
for $\tau > 3$ there are more favorable situations for D agents and 
disfavorable for C agents, the mean cooperation of the system when the 
dynamical equilibrium is reached, $c_\infty=\frac{1}{N_{ag}} 
\sum_{{N_{ag}}}\, c(x,y,t)$ -after the transient-, will be 
smaller than when $\tau < 3$.

Table 2 summarizes the utilities of a player for each possible
configuraqtion of his neighbors for the case of Moore neighborhood.
 
\vskip 1mm

\begin{center}

\begin{tabular}{|c||c|c|c|c|c|c|c|c|c|}  \hline
  & 8C, 0D & 7C, 1D & 6C, 2D & 5C, 3D & 4C, 4D & 3C, 5D & 2C, 6D & 1C, 7D & 0C,8D \\ \hline \hline   
C & 8 & 7-$\tau$  & 6-2$\tau$ & 5-3$\tau$ & 4-4$\tau$ & 3-5$\tau$  & 2-6$\tau$ &
1-7$\tau$ & -8$\tau$  \\ \hline 
D   & 8$\tau$ & 7$\tau$ -1  & 6$\tau$-2 & 5$\tau$-3 & 4$\tau$-4 & 3$\tau$ -5  &
2$\tau$-6 & $\tau$-7 & -8  \\ \hline
\end{tabular}

\end{center}

\vskip 1mm

Table 2. The same as Table 1 but for 
Moore neighborhood.

\vspace{1mm}

A completely analogous reasoning for the Moore neighborhood leads 
to three ''critical" values: $\tau^*_1= 5/3$,  $\tau^*_2= 3$ and 
$\tau^*_3= 7$. 
Here we would expect also that $c_\infty$ will diminish as $\tau$ crosses
each frontier value $\tau^*_i$ from left to right.

\section{RESULTS}

To avoid dependence on the initial conditions the measures correspond to
averages taken over an ensemble of 100 systems with arbitrary
initial conditions. 
In general, the results for the asymptotic regime, after a transient,
become almost
independent of the lattice size $L$ for $L\gta 100$. Therefore in what
follows, unless it is stated otherwise, the results correspond to 
simulations performed in $100\times100$ lattices.

As we have anticipated, we observe that the stationary state of the system 
changes as the parameter $\tau$ moves from one region to another (two regions 
in the case of $z=4$ von Neumann neighborhood and four regions for $z=8$ 
Moore neighborhood).

\subsection{Asymptotic average fraction of cooperators $\bar{c}_\infty$}

The asymptotic or equilibrium mean fraction 
of C-agents $\bar{c}_\infty$\footnote{The upper bar in $\bar{c}_\infty$
denote an average over 100 simulations with different initial conditions.}, 
takes constant values in each of the 
regions delimited by the ''critical" $\tau^*$. 
Hence we have one sharp step at $c_\infty=3$ for $z=4$ and 
three sharp steps at $c_\infty=\frac{5}{3}$, 3 $\&$ 7 for $z=8$.

It is interesting to compare the $\bar{c}_\infty$,  produced by simulations,
with the $c^{MF}_\infty$  obtained by elementary
calculus using a Mean Field (MF) approximation that neglects all spatial
correlations (see APPENDIX I).

In Tables 3 and 4 we present the $\bar{c}_\infty$ and  $c^{MF}_\infty$ for
$z=4$ and $z=8$ respectively.
Clearly, as expected, the MF approximation improves increasing $z$. 
In addition, divergences between spatial games and the MF approximation become
maximum in the "cooperative" sector of the parameter $\tau$
(leftist region, producing $c_\infty \gta 0.5$ ). This
can be explained in terms of the particular cluster structure of that region 
exhibiting power law scalings (see next subsection).

\vskip 1mm

\begin{center}

\begin{tabular}{|c|c|c|}  \hline
z=4  & Simulations & MF \\ \hline \hline   
$\tau < 3$   & $0.485 \pm 0.002$  &  0.430 \\ \hline 
$\tau \ge 3$  & $0.280 \pm 0.002$ &  0.342  \\ \hline
\end{tabular}

\vskip 1mm

Table 3. The asymptotic fraction of cooperators $c_\infty$ for $z=4$
von Neumann neighborhood.
Column 2: simulations. Column 3: MF approximation (see APPENDIX I).

\end{center}

%\begin{equation}
%\label{eq:ceq4v}
%\quad 0.485 \pm 0.002 \quad \mbox{if} \quad \tau <3 \\
%\quad 0.280 \pm 0.002 \quad \mbox{if} \quad \tau >3 
%\end{array} \right.
%\end{equation} 
%{\it i.e.} a sharp step at $\bar{c}_\infty=3$.
%For the Moore neighborhood we have: 
%\begin{equation}
%\label{eq:ceq8v}
%\bar{c}_\infty = \left\lbrace \begin{array}{l}
%\quad 0.563 \pm 0.002 \quad \mbox{if} \quad \tau \in (1,\frac{5}{3}) \\
%\quad 0.436 \pm 0.002 \quad \mbox{if} \quad \tau \in (\frac{5}{3},3) \\
%\quad 0.366 \pm 0.003 \quad \mbox{if} \quad \tau \in (3,7) \\ 
%\quad 0.320 \pm 0.003 \quad \mbox{if} \quad \tau >7
%\end{array} \right.
%\end{equation}     

\begin{center}

\begin{tabular}{|c|c|c|}  \hline
z=8  & Simulations & MF \\ \hline \hline
$1< \tau < 5/3$   & $0.563 \pm 0.002$  &  0.461 \\ \hline
$5/3 \le \tau < 3$  & $0.436 \pm 0.002$ &  0.420  \\ \hline
$3 \le \tau < 7$   & $0.366 \pm 0.003$  &  0.386 \\ \hline
$8 \le \tau$  & $0.320 \pm 0.003$ &  0.334  \\ \hline  
\end{tabular}

\vskip 1mm

Table 4. The asymptotic fraction of cooperators $c_\infty$ for $z=8$
More neighborhood.
Column 2: simulations. Column 3: MF approximation (see APPENDIX I).

\end{center}

\subsection{Spatial Patterns: The Cluster Structure}

{\it Von Neumann neighborhood}

\vspace{1mm}

\begin{center}
\begin{figure}[h]
\centering
\psfig{figure=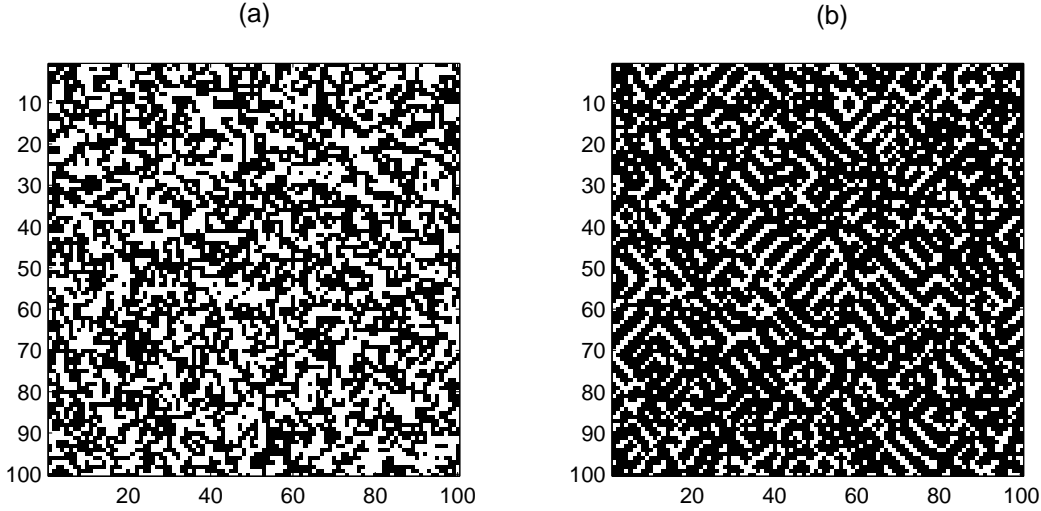,height=7cm}
\caption{Asymptotic ''cooperation maps" 
for: \textbf{(a)} $\tau <3$, \textbf{(b)} $\tau >3$. 
Black=D, white=C.}
\label{fig:coopmap4vT2andT4}
\end{figure}
\end{center}

In Fig. \ref{fig:coopmap4vT2andT4} we present 
snapshots -after the transient- of the cellular automaton for
$\tau < 3$ and $\tau > 3$.
These ''cooperation maps" illustrate the differences between the
typical spatial patterns that arise in the two parameter regions
divided by $\tau*=3$.

\vspace{-9mm}

For $\tau<3$ we found that: 

I) Although the asymptotic probability for D agents is $\bar{d}_\infty=
1-\bar{c}_\infty\simeq 0.5$,
which is below the percolation threshold $p_c\approx0.59275$, 
giant spanning D clusters often occur.
Percolation below threshold is a known fact in other models.
In general, when there are 
correlations between the sites, the threshold is shifted. As it happens, 
for instance, in the square Ising model percolation occurs, at the 
critical temperature, when the concentration is also 0.5. \\

II) Different quantities behave as power laws
implying thus the emergence of scale free phenomena.
For instance, the size distribution of clusters of D agents 
exhibits power law scaling. 

\vspace{1mm}

For $\tau >3$  the distribution of D clusters is bimodal with a
peak for very small clusters (size=1) and a secondary peak for very large
clusters. The main peak for very small clusters can be explained by the 
small correlation length. On the other hand, the secondary peak for very 
large sizes arises because the probability of a given site to be in the D 
state $\bar{d}_\infty \equiv 1 -\bar{c}_\infty$ is over the site percolation 
threshold and thus spanning clusters are much more abundant than when 
$\tau<3$ in which case $\bar{d}_\infty < p_c$.  

\begin{center}
\begin{figure}[h]
\centering \psfig{figure=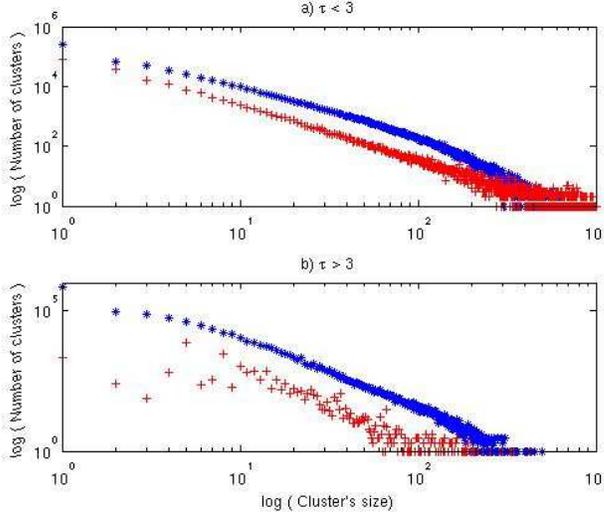,height=7cm}
\caption{Number of clusters of C (*) and D (+) agents vs. size of the 
clusters for the von Neumann neighborhood in a 
400$\times$400 lattice. 
The clusters are summed over the last 150 lattice sweeps after the transient for: \textbf{(a)} 
$\tau <3$, \textbf{(b)}  $\tau >3$. In both cases giant spanning D clusters 
were not included}
\label{fig:clusCD300orig}
\end{figure}
\end{center}

Fig. \ref{fig:clusCD300orig} 
is a plot of the log of the number of clusters of C and D agents 
vs. the log of their size for 
$\tau <3 $ and $\tau >3$ using $400\times 400$ lattices.
In both cases giant spanning clusters of D agents were excluded. 
This, in particular for $\tau > 3$, eliminates a large number of
clusters belonging to the secondary peak of its bimodal distribution
and explains why there are less "+" points in Fig.\ref{fig:clusCD300orig}-
\textbf{(b)} than in \ref{fig:clusCD300orig}-\textbf{(a)} (the shortage
of "*" points, representing C clusters, obviously is related to the fact 
that $c_\infty$ is smaller on the $\tau >3$ side).

%\vspace{-12 mm}

The data points for D clusters seem consistent with a power law scaling 
over a couple of decades,
with a critical exponent of approximately $-1.79 \pm 0.02$. The graphic
also shows a difference between C and D clusters: the first ones exhibit 
much greater deviations from an exact power law although they also occur 
over a wide range of scales. 
This asymmetry can be traced to 
the difference that exists for the possible stable configurations of 
clusters of C's or D's; while the first ones need at least three 
C neighbors to remain C, the second ones can do well with only two 
C neighbors. Then the D agents can form thinner clusters than  
the C agents. This fact increases the probability of agents D to yield 
larger clusters. This also can explain why 
although the equilibrium probability for D agents is below the percolation 
threshold, giant spanning D clusters are observed.

For $\tau >3$ the situation changes drastically as Fig. 
\ref{fig:clusCD300orig}.(b) reflects, here it can be 
seen that the data don't fit well with a power law neither for D nor for C
clusters. \\

{\it Remark} -
To check that the power law scaling is not dependent on the  
particular parameterization of the payoff matrix we are using, 
we measured the cluster distribution for many other payoff matrices not
described by (\ref{eq:payoffmatrix}). For instance, we considered 
this alternative parameterization of the
payoff matrix
\begin{equation}                                                                    
\nonumber {\mbox M'}=\left(\matrix{
(1,1)&(\tau/2-3 ,\tau)\cr (\tau,\tau/2-3)&(-1,-1) \cr}\right), 
\label{eq:param2}
\end{equation}                                                                                
with $3-\tau/2 < -1 < 1 < \tau $ 
Again, we found power law behaviour for the 
leftist region in $\tau$. Thus, it seems that this power law scaling for an
entire collection of PD payoff matrices is a robust property of the model.

\vspace{2mm}

\begin{figure}
   \centering 
   \psfig{figure=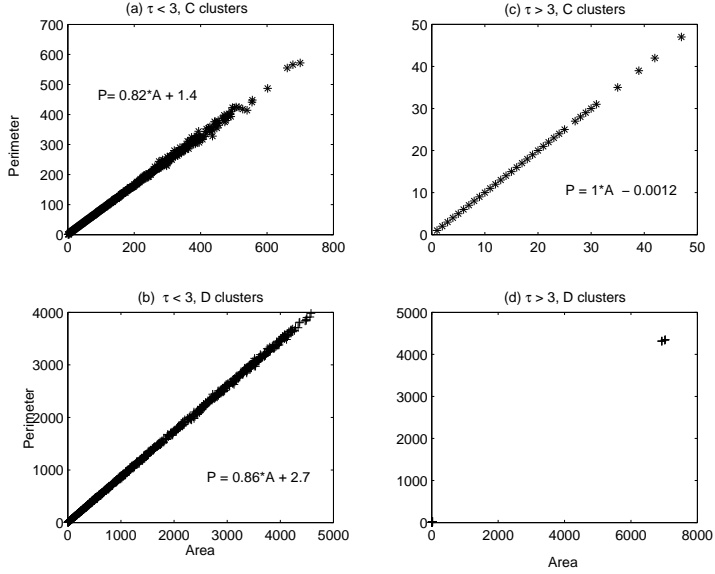,height=8cm}
   \caption{Perimeter vs. area of the clusters of C and D agents for z=4.
           The perimeter's values plotted are averages of perimeters of  
           clusters of the same size, taken over the last 500 lattice sweeps
           after the transient.}
   \label{fig:perimvsarea}
\end{figure}

Another clue about the dynamics of the clusters can be obtained by 
examining the relation of the perimeter to the area of the clusters. 
We define the perimeter of a cluster C (D) as the set of sites $(x,y)$
with behavioral variable $c(x,y)=1$ ($c(x,y)=0$)
belonging to the cluster with at least one neighbor with the opposite 
behavioral variable i.e. $c(x,y)=0$ ($c(x,y)=1$). The mean
perimeter $P(A)$, for a given area $A$, is then given by averaging over 
all the perimeters of clusters with area $A$.
Fig.\ref{fig:perimvsarea} shows that for $\tau<3$ the perimeter scales 
linearly with the area, that is, at the fastest rate possible, 
implying that the clusters are highly ramified. The fraction of the area that 
is interior to the clusters can be easily calculated.

%\vspace{-5mm}

By fitting the point of Figs.\ref{fig:perimvsarea}.(a) and 
(b) we get the following expressions for the perimeter as a function 
of the area, for $\tau < 3$:
\begin{equation}
\begin{array}{ll}

P_{C}\approx0.82 A_{C} & \quad \mbox{for clusters of C agents} \\
P_{D}\approx0.86 A_{D} & \quad \mbox{for clusters of D agents}.

\end{array}
\end{equation}

Then the cluster interior fraction is $F=\frac{A-P}{A}$. Thus we get 
that approximately:
\begin{equation}
\begin{array}{ll}

F_{C} \simeq 0.18 & \quad \mbox{for clusters of C agents} \\
F_{D} \simeq 0.16 & \quad \mbox{for clusters of D agents}

\end{array}
\end{equation}  

This shows that the clusters have almost no interior, and confirms our 
previous observation concerning that the clusters of D agents are thinner 
than those of C agents. This supports quantitatively the explanation of why
percolation of D agents is observed but no of C agents.
The linear behavior shown in Fig.\ref{fig:perimvsarea}.(c), 
which slope approximatelly equal to 1, can be understood by inspection of 
Fig.\ref{fig:coopmap4vT2andT4}.(b) where is clearly seen that 
the C agents form small "laddered" clusters in which the perimeter is equal 
to the area.

\vspace{2mm}

{\it Moore neighborhood}
 
\vspace{1mm}

\begin{figure}
    \psfig{figure=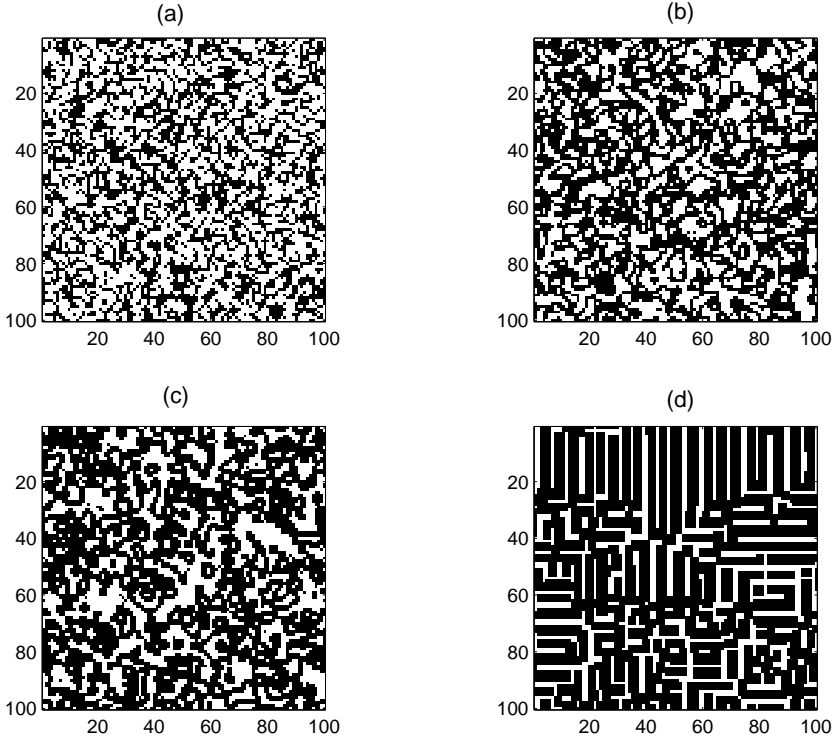,height=10cm}
   \caption{Cooperation maps for Moore neighborhood at equilibrium 
(after $10^{5}$ iterations) for: 
\textbf{(a)} $\tau \in (1,\frac{5}{3})$, \textbf{(b) 
$\tau \in (\frac{5}{3},3)$}, \textbf{(c)} $\tau \in (3,7)$ and 
\textbf{(d)} $\tau >7$. Black=D, white=C.}
   \label{fig:coopmap8v}
\end{figure}

For arbitrary random initial conditions, the equilibrium cooperation maps 
are shown in Fig.\ref{fig:coopmap8v} for $\tau$ 
in the different regions of interest.

\nopagebreak

As it can be seen from Table 4, when $\tau$ is within the interval 
$(1,\frac{5}{3})$, $\bar{c}_\infty \simeq 0.6$ which is higher than the 
values obtained for the von Neumann neighborhood for any $\tau$. 
This implies that 
increasing the number of neighbors in general produces a higher fraction of  
cooperators, although this higher value of 
$\bar{c}_\infty$ is stable for narrower domain values of $\tau$. 
We checked this for the case in which 12 neighbors are taken into account, 
achieving a value of $\bar{c}_\infty \simeq 0.8$ for $\tau \in (1,\frac{7}{5})$.
\\

\begin{figure}
   \centering
   \psfig{figure=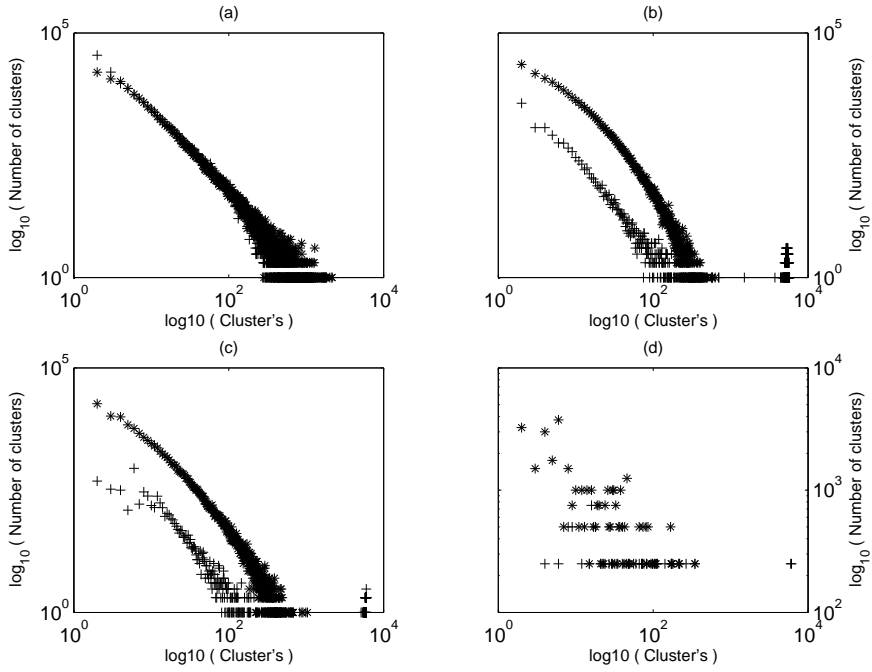,height=9cm}
   \caption{Number of clusters of C(*) and D(+) agents vs. size of the clusters, summed over the last 500 times after $10^{4}$ iterations for $z=8$, in logarithmic scale. The plots correspond to: \textbf{(a)} $\tau \in (1,\frac{5}{3})$, \textbf{(b)} $\tau \in (\frac{5}{3},3)$, \textbf{(c)} $\tau \in (3,7)$, \textbf{(d)} $\tau > 7$. There is a percolation peak for clusters of D agents in \textbf{(b)}, \textbf{(c)} and \textbf{(d)} since they are above the percolation threshold ($d > p_{c}$).}
   \label{fig:clustersCD8v}
\end{figure}

Let us analyze what happens to the clusters of C's and D's 
for the different values of $\tau$, this time for the Moore neighborhood. 
The results are shown in Fig. 
\ref{fig:clustersCD8v}.

%\nopagebreak

In Fig.\ref{fig:clustersCD8v}.(a), corresponding to 
$\tau \in (1,\frac{5}{3})$ and $c_\infty \simeq 0.57$, we can observe 
power law behavior for both clusters of C and D agents, with the same 
critical exponent of approximately $-1.62 \pm 0.02$. This symmetry 
between C's and D's is broken when we take $\tau \in (\frac{5}{3},3)$ 
(Fig.\ref{fig:clustersCD8v}.(b), $c_\infty \simeq 0.44$): here we 
recover the kind of behavior we found for $\tau <3$ in the case of the von 
Neumann neighborhood (see Fig. \ref{fig:clusCD300orig}.(a)), 
for which the power law scaling for D agents is much more claear than 
for C agents. 
In this case we find an exponent of approximately $-1.98 
\pm 0.04$. Remarkably, criticality seems to persist, although not so 
clearly as in the previous cases, even for values of $\tau$ in the 
interval $(3,7)$  (Fig.\ref{fig:clustersCD8v}.(c)). 
For $\tau >7$, power law behavior is completely lost, as 
Fig.\ref{fig:clustersCD8v}.(d) shows.\\

\subsection{Power Spectrums}

The power laws we found for {\it spatial} observables might be interpreted as 
signatures of self-organized criticality 
(SOC). In order to elucidate the 
criticality or not of the dynamics we analyzed {\it temporal} correlations.
Specifically, we calculated the
power spectrum $P(f)$ ({\it i.e.} the absolute value of the Fourier
transform) of the time autocorrelation function $G(t)$ of the cooperative 
fraction $c(t)$. 
$G(t)$ is defined as:
\begin{equation}
G(t) \equiv <c(t_0)c(t_0+t)>-<c(t_0)>^2,
\label{eq:G}
\end{equation}
where the average is taken over all possible temporal origins $t_0$.

\begin{figure}
   \centering
   \psfig{figure=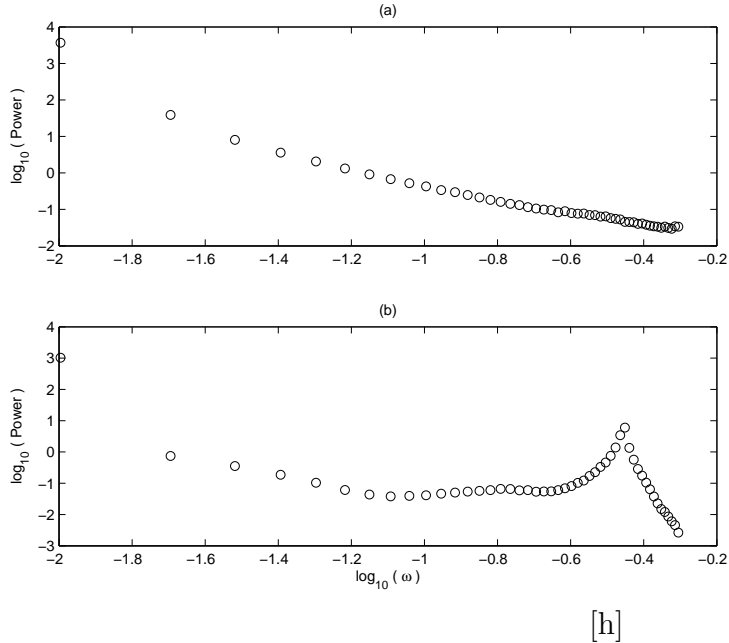,height=8cm}[h]
   \caption{Power spectrum for $z=4$ Von Neumann neighborhood
    : \textbf{(a)} $\tau <3$, \textbf{(b)} $\tau >3$.}
   \label{fig:specT2and4}
\end{figure}

It turns out that although the transients are not very long, $P(f)$ exhibits 
power law behavior, for the same range of values of $\tau$ we found this type
of behavior for the cluster size distributions, for almost two decades. 
For instance, in the case of the von Neumann neighborhood,
we have a power law power spectrum for $\tau <3$ which is lost for $\tau >3$
(which is consistent with the fact that the simulations have shown that 
for this region the system behaves periodically, with a very short period). 
This is shown in Fig.\ref{fig:specT2and4}. 

The correlation function $G(t)$ 
is calculated for the transient. In order to maximize this transient  
an initial $c(t=0)=0.1$ very different from the known
equilibrium value of $c_\infty\simeq0.5$ was taken together with a
large lattice of $1000 \times 1000$.
This power law scaling of $P(f)$, for the same region we found this type
of behavior for the cluster sizes, can be interpreted as another signature for the possible existence of critical dynamics.

\subsection{Asynchronous dynamics}

As we mentioned in the previous section, besides exploring the synchronous 
dynamics, we also performed some runs using the asynchronous dynamics, in 
which the state of each agent is updated after he played with his
neighborhood.

The asynchronous update produce a much less interesting situation.
The power laws are lost, both for the von
Neumann and Moore neighborhoods: we find no power laws for the cluster 
sizes nor for the power spectrum and the cooperation values decrease 
significantly. Still, there is a change in the mean value of the 
cooperation as the parameter $\tau$ goes through the critical values 
calculated earlier. For the \textit{von Neumann} neighborhood, 
for $\tau<3$, $\bar{c}_\infty \simeq 0.34$. 
For $\tau>3$ cooperation decreases to 
$\bar{c}_\infty \simeq 0.23$ and there is no clear pattern of behavior. For 
the \textit{Moore} neighborhood results are similar, with $\bar{c}_\infty 
\simeq 0.34$, $0.30$, $0.21$ and $0.13$ for $\tau \in (1,\frac{5}{3})$, 
$(\frac{5}{3},3)$, $(3,7)$ and $\tau >7$ respectively.

\section{CONCLUSIONS}
 
For a cellular automata, representing a system of agents playing the IPD 
governed by Pavlovian strategies in a simple territorial setting, we 
explored its steady states for different values of the parameter $\tau$
, which measures the ratio of \textit{temptation to defect} to 
\textit{reward}. Both for the Von Neumann and Moore neighborhoods we found
sharp steps for $\bar{c}_\infty$ vs. $\tau$ (one step in the first case and
three steps in the second case).

We found power-law scaling for different quantities, measuring
either spatial (cluster size distributions) or temporal correlation ($P(f)$), 
for {\it entire} regions in parameter $\tau$ space.  
All this may be interpreted as consistent 
evidences of self-organized criticality in a spatial game which is not 
evolutionary (at least in the ordinary Darwinian sense). 
This result, which is qualitatively robust 
against changes of the payoff matrix and the neighborhood, 
is novel (as far as we know). 
[It is worth to mention that the parameterization (\ref{eq:param2}) allows to
study two other games besides the PD: If $-1 < \tau < 1$ ($R>T>P>S$)
the game is known as ''Stag Hunt" (SH) while when $4<\tau<8$ ($T>R>S>P$)
the game is called the ''Hawk-Dove" (H-D). 
We simulated these two games, which are popular in Social
Sciences and Biology respectively, and, in contrast to what happen with the
PD, we found no power law behavior \cite{future}.]
On the other hand, the occurrence of critical dynamics in certain spatial 
evolutionary games has been observed. For instance, in ref. \cite{kd98}
it was shown that for certain range of a parameter, which determines the
punishment, the spatial HD game exhibits large temporal and spatial
correlations and various processes governed by power-laws. This is in
contrast with the simplified version of the PD considered in ref. \cite{nm92}, 
which does not exhibit complex critical dynamics of this type, rather it has 
periodic or chaotic dynamics. Nevertheless, for a stochastic version of this 
evolutionary weak dilemma, power law behavior consistent with directed
percolation has been measured \cite{st98}.

We also have shown that percolation below the threshold value occurs for 
D-agents for the case of the von Neumann neighborhood. 
The asymmetry between C and D clusters, even in cases in which both types of
agents appear with equal probability, can be explained in terms of the
Pavlovian strategy and the asymmetry of the payoffs (see Table 1).   

A result worth remarking is that the degree of cooperation can be increased 
by enlarging the neighborhood but, simultaneously, the temptation parameter 
 $\tau$ must be restricted to smaller values.

Another interesting general result is the effect of changing the
dynamics from synchronous to asynchronous. The scale invariance we found 
for the synchronous update
disappear when we turn to the asynchronous update. 
The fact that the general qualitative behavior of asynchronous models 
may differ greatly from that of the synchronous version was noticed in  
\cite{hg93}. 

\begin{center}  
\begin{figure}[h]
\centering \psfig{figure=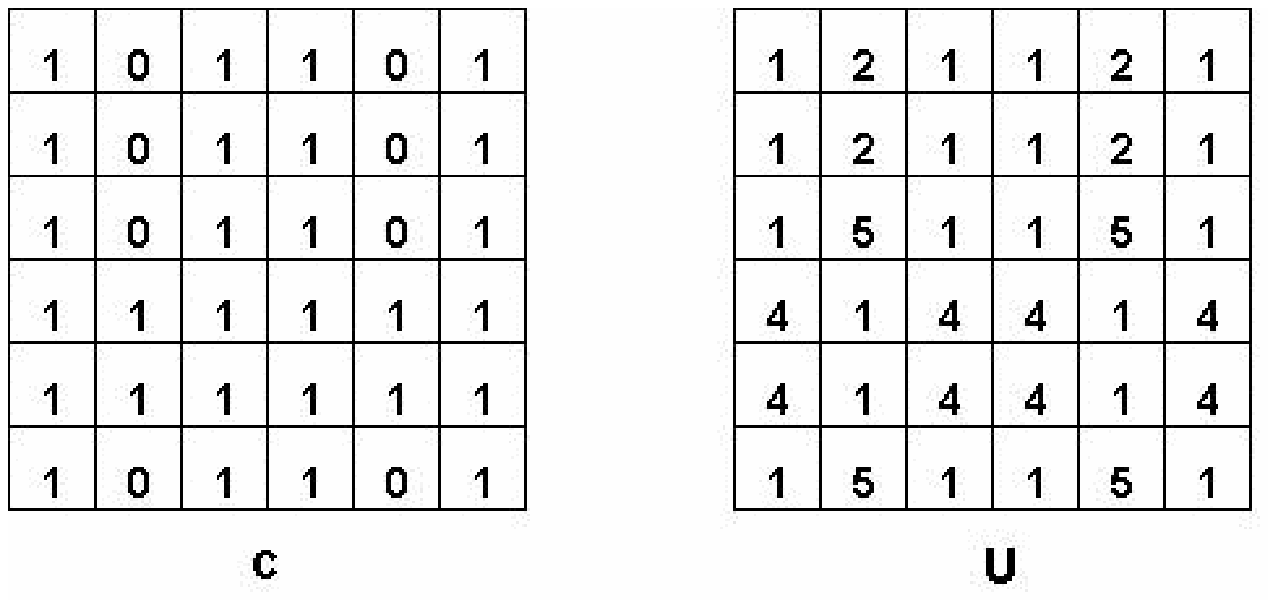,height=6cm}
\caption{{\it Pareto Optimal states} configuration for a small 
$6 \times 6$ lattice, $z=4$ and $\tau =3$. 
Left: the $c(i,j)$ matrix.
Right: the corresponding utilities $U(i,j)$: the utilities for all
the agents are positive and thus they don't change their behavioral
variables.}
\label{fig:IW}  
\end{figure}
\end{center}

Let us mention some interesting future extensions of the work
presented here.
For instance, we observed that for small lattices this simple deterministic 
system often reaches true equilibrium configurations, in which all the 
agents are happy (all get utilities above 0) and do not change their
respective states. In other words,  
{\it Pareto Optimal} states ($POS$) 
{\it i.e.} states in which none of the players 
can increase their payoff without decreasing the payoff of at least one of the
other players. 
In Fig.\ref{fig:IW} an example of such equilibrium states 
is presented for a small ($6 \times 6$) lattice, $z=4$ and $\tau=2$.

When the lattice size grows the system becomes unable to
reach these $POS$. The explanation we found for this is 
, as the size grows, the fraction of $POS$ with respect to the possible 
configurations decreases. Additionally, it is plausible that
the entirely deterministic update does not provide a path in 
configuration space connecting the initial state with an $POS$. 
The introduction of noise in the update rule, in some particular cases,
might help promoting ergodicity.
The effect of the introduction of noise in spatial evolutionary games was
analyzed for example in \cite{ln94} and \cite{nbm94}.
An interesting goal is how to use noise to avoid entrainment in non efficient
states {\it i.e.} to implement a sort of {\it simulated annealing} approach 
\cite{kgv83} allowing to reach these optimal equilibriums.

Another issue that seems worth exploring is the extension of the present
approach, beyond the PD game, to games that are useful to model other 
different everyday situations, like the "Stag Hunt", "Chicken", etc
\cite{future}.  
  
Finally, after we concluded this manuscript, one of the referees pointed out
the study of the PD game of Posch {\it et al} \cite{pps99} 
using "win-stay, lose-shift" strategies in a non spatial setup.
This work offers an stimulating discussion of when can satisficing become  
optimizing.
 
\vspace{2mm}
 
{\bf Acknowledgments.}
 
We are grateful to Professor Dietrich Stauffer for useful comments on
percolation. S.V. wish to thank D.Guerra for sharing his programming 
skills.
 
\vspace{1.5cm}

{\bf APPENDIX I: MEAN FIELD COMPUTATIONS}

\vspace{2mm}

Estimate of $c_\infty$ can be obtained by elementary
calculus using a Mean Field approximation that neglects all spatial
correlations.

Once the stationary state was reached, the transitions from D to C, on average,
must equal those from C to D.  
Thus, the average probability of cooperation $c_\infty$ is obtained by
equalizing the flux from C to D, $J_{CD}$, to the flux from D to C,
$J_{DC}$.
The possible utilities for a C player range from $R=z\times1=z$ to 
$S=-z\tau$ (see Table 1 and Table 2).  
Let us consider by separate the $z=4$ von Neumann neighborhood and the 
$z=8$ Moore neighborhood.

\vspace{2mm}

$z=4$

\vspace{1mm}

We have two different situations depending on the
value of $\tau$: $\tau < 3$ or $\tau \ge 3$. 

\begin{itemize}

\item $\tau < 3$: 

In that case, the utilities $U_C$ ($U_D$) of a C (D) player are 
negative, and thus he changes from C to D (D to C) 
if at least 2 (3) neighbors play D.
For a given average probability of cooperation $c$, the probabilities of
a C agent facing 2, 3 and 4 neighbors playing D are respectively: 
$c^3(1-c)^2$, $c^2(1-c)^3$ and $c(1-c)^4$. 
Consequently, $J_{CD}$ can be written as:
\begin{equation}
J_{CD} \propto c^3(1-c)^2+c^2(1-c)^3+c(1-c)^4.
\label{eq:JCD}
\end{equation}
On the other hand, 
the probabilities of a D agent facing 3 and 4 neighbors playing D are   
respectively: $(1-c)^4c$ and $(1-c)^5$. Therefore $J_{DC}$ is given by:  
\begin{equation}
J_{DC} \propto c(1-c)^4+(1-c)^5.
\label{eq:JDC}
\end{equation}
Thus the algebraic equation for $c_\infty$ is:
\begin{equation}
c_\infty^3+c_\infty^2(1-c_\infty)-(1-c_\infty)^3=0,
\label{eq:cz4Tlessthan3}
\end{equation}
with only one real root in the interval [0,1]: $c^{MF}_\infty=0.430$.

\item $\tau \ge 3$:

In that case, the utilities $U_C$ ($U_D$) of a C (D) player are
negative, and thus he changes from C to D (D to C)
exept (only) if he has all his 4 neighbors playing C (D).
Therefore, $J_{CD}$ must be modified summing a term
$c^4(1-c)$ to eq. (\ref{eq:JCD}) and the term $c(1-c)^4$ must be supressed 
from the expresion (\ref{eq:JDC}) for $J_{DC}$.
Hence, we get the following algebraic equation for $c_\infty$:
\begin{equation}
c_\infty^4+c_\infty^3(1-c_\infty)+c_\infty^2(1-c_\infty)^2-c_\infty(1-c_\infty)^3-
(1-c_\infty)^4=0, 
\label{eq:cz4Tgreaterthan3}
\end{equation}
with only one real root in the interval [0,1]: $c^{MF}_\infty=0.342$.

\vspace{2mm}

$z=8$

\vspace{1mm}

We have four different situations depending on the region in the parameter 
space $\tau$. The corresponding polynomials for $c_\infty$ 
are obtained exactly
as it was done for $z=4$ and one can easely check that are given by:

\item $1 < \tau < 5/3$
\begin{equation}
c_\infty^5+c_\infty^4(1-c_\infty)-(1-c_\infty)^5=0,
\label{eq:cz8region1}
\end{equation}
with only one real root in the interval [0,1]: $c^{MF}_{eq1}=0.461$.

\item $5/3 \le \tau < 3$
\begin{equation}
c_\infty^6+c_\infty^5(1-c_\infty)+c_\infty^4(1-c_\infty)^2+
c_\infty^3(1-c_\infty)^3-(1-c_\infty)^6=0,                    
\label{eq:cz8region2}      
\end{equation}
with only one real root in the interval [0,1]: $c^{MF}_{eq2}=0.420$.
 
\item $3 \le \tau < 7$
\begin{eqnarray}
c_\infty^7+c_\infty^6(1-c_\infty)+c_\infty^5(1-c_\infty)^2+ c_\infty^4(1-c_\infty)^3+
\nonumber \\ 
c_\infty^3(1-c_\infty)^4+c_\infty^2(1-c_\infty)^5-(1-c_\infty)^7=0,
\label{eq:cz8region3}      
\end{eqnarray}
with only one real root in the interval [0,1]: $c^{MF}_{eq3}=0.386$.

\item $7 \le \tau$                                 
\begin{eqnarray}                                       
c_\infty^8+c_\infty^7(1-c_\infty)+c_\infty^6(1-c_\infty)^2+c_\infty^5(1-c_\infty)^3+
c_\infty^4(1-c_\infty)^4+ \nonumber \\
c_\infty^3(1-c_\infty)^5+ 
c_\infty^2(1-c_\infty)^6+c_\infty(1-c_\infty)^7-(1-c_\infty)^8=0,
\label{eq:cz8region4}                                  
\end{eqnarray}                                         
with only one real root in the interval [0,1]: $c^{MF}_{eq4}=0.334$.

\end{itemize}


\begin{thebibliography}{99}
\bibitem{future} Results for SH and H-D will be published elsewhere.
\end{thebibliography}
\end{document}